# Direct visualization of the magnetostructural phase transition in nano-scale FeRh thin films using differential phase contrast imaging

Trevor P. Almeida[1]*, Damien McGrouther[1], Rowan Temple[2], Jamie Massey[2], Yue Li[1], Thomas Moore[2], Christopher H. Marrows[2], Stephen McVitie[1]

[1] SUPA, School of Physics and Astronomy, University of Glasgow, Glasgow, G12 8QQ, UK.
[2] School of Physics and Astronomy, University of Leeds, Leeds LS2 9JT, UK.

**Abstract**

To advance the use of thermally-activated magnetic materials in device applications it is necessary to examine their behaviour on the localised scale *in operando* conditions. Equi-atomic FeRh undergoes a magnetostructural transition from an antiferromagnetic (AF) to a ferromagnetic (FM) phase above room temperature (~ 75 to 105 °C) and hence is considered a very desirable material for the next generation of novel nanomagnetic or spintronic devices. For this to be realised, we must fully understand the intricate details of AF to FM transition and associated FM domain growth on the scale of their operation. Here we combine *in-situ* heating with a comprehensive suite of advanced transmission electron microscopy techniques to investigate directly the magnetostructural transition in nano-scale FeRh thin films. Differential phase contrast imaging visualizes the stages of FM domain growth in both cross-sectional and planar FeRh thin films as a function of temperature. Small surface FM signals are also detected due to interfacial strain with the MgO substrate and Fe deficiency after HF etching of the substrate, providing a directional bias for FM domain growth. Our work provides high resolution imaging and quantitative measurements throughout the transition, which were previously inaccessible, and offers new fundamental insight into their potential use in magnetic devices.

**\* Correspondence to: trevor.almeida@glasgow.ac.uk**

**Introduction**

The ordered $\alpha''$ alloy of iron-rhodium ($Fe_{48}Rh_{52}$ to $Fe_{56}Rh_{44}$) has gained significant interest due to its magnetostructural transition from its antiferromagnetic (AF) to ferromagnetic (FM) phase[1]. This equi-atomic, CsCl-structured $\alpha''$ alloy undergoes a first-order transition from its room-temperature AF state to FM state between ~ 75 to 105 °C, which is accompanied by a 1% lattice volume expansion[2]. Hence, nano-scale thin films of FeRh can present AF / FM phase co-existence and hysteresis, where the transition temperature and associated thermal hysteresis depend on the film thickness[3], lattice strain behaviour and additional chemical doping[4]. At intermediate stages during the transition, the co-exisiting AF / FM phase regions dynamically evolve or disintegrate with temperature and are separated by phase boundary (PB) domain walls (DWs). Exploitation of the PB motion and its effective control is considered very desirable for the next generation of spintronic devices. For example, the PBs can be systematically driven by heating FeRh films grown with differential gradients of elemental Ir- and Pd-doping, as determined by a corresponding change in resistivity[5]. However, the dynamical FM domain nucleation, growth and coalescence stages on heating, and subsequent separation, disintegration and annihiation stages on cooling, as well as the PB behaviour, are not accessible by bulk magnetic measurements.

Imaging techniques sensitive to magnetic structure and with spatial resolution in the 10's of nanometres range, including magnetic force microscopy[6,7], X-ray magnetic circular dichroism (XMCD)[7,8] and scanning electron microscopy with polarization analysis (SEMPA)[9] have been used to observe the phase coexistence in FeRh thin films, showing the first order transition from the nucleation of domains regime to be distinct from the domain growth regime[8,10]. Further, X-ray photoemission microscopy in XMCD has revealed the effects of lateral confinement of FeRh islands on the transition temperature, resulting in a 20 K variation in transition temperature in small islands, and showed that their ion-beam-damaged edges act as favourable nucleation sites[11]. Nevertheless, these techniques are typically limited to a spatial resolution of ~ 20 - 30 nm and penetration depth of a few nm[12,13]. Hence, in order to elucidate fully the localised and dynamic domain evolution / dissipation throughout the magnetostructural transition with sufficient detail, it is necessary to investigate the thermally-induced domain growth at the highest spatial resolution during *in situ* heating.

Aberration-corrected transmission electron microscopy (TEM) techniques are well known for enabling the imaging of both physical and chemical structure of sufficiently thin, electron transparent, samples with atomic-scale spatial resolution. High spatial resolution magnetic imaging can also be performed utilising a family of techniques collectively known as Lorentz microscopy. These include Fresnel imaging[14-16], off-axis electron holography[17-20] and differential phase contrast (DPC) imaging[16,21,22]. Recent advancements of modern aberration-corrected TEMs have also improved the spatial resolution of magnetic imaging to approach ~ 1 nm[23]. Fresnel imaging has revealed the FM domain structure in high quality 2D planar-view FeRh thin films produced for magnetic imaging within the TEM[24]. Electron holography has also exposed an inhomogeneous spatial distribution of the transition temperature along the growth direction in cross-sectional FeRh thin films, as well as a regular spacing of the nucleated FM domains[25]. However, preparation of cross-sectional thin film TEM lamellae is inherently destructive and shape anistropy dominates magnetically compared to its continuous film state. Accordingly, conventional DPC imaging of planar FeRh thin films has provided quantitative measurements from individual DWs as a function of temperature, including a general overview of the nucleation, growth and coalescence stages during the transition[26]. Yet this type of conventional DPC imaging is susceptible to unwanted signal variation due to strong diffraction contrast from grain boundaries and crystallinity, as it employs segmented detectors to measure differential signals that relate to in-plane magnetism but which can also arise from crystallographic directional scattering. To overcome this, the recent advent of fast direct electron pixelated detectors has revolutionised the ability to acquire large mount of images in relatively short time periods. DPC type imaging is performed by capturing images of the transmitted electron disc for every electron beam scan location with advanced data processing, based on disc edge detection, enabling isolation of small Lorentz deflections from directional artefacts introduced by diffraction contrast[27]. This technique belongs to a wider, relatively new, family of techniques termed "4D STEM". Thus, when reporting results from both DPC techniques we refer to these as being from segmented detector DPC or 4D STEM DPC.

Herein, this study employs a comprehensive suite of advanced magnetic TEM imaging and scattering techniques to investigate the magnetostructural transition in cross-sectional and planar FeRh thin films, as a function of temperature. Conventional TEM characterisation confirms their chemical and structural properties, whilst segmented detector DPC imaging reveals the origin of a small FM signal at room temperature and FM domain growth of cross-sectional FeRh films during heating. Localised insight into the AF to FM domain evolution within a planar FeRh film, and subsequent reversal, is provided by three complementary Lorentz microscopy techniques to investigate the nature of the appearance/disappearance of the FM phase. Defocused Fresnel imaging in TEM mode was utilised to obtain high contrast, real space imaging of the phase transition. Small angle electron scattering (SAES)[28] provided a counterpart quantitative, reciprocal space, analysis of the same regions imaged by Fresnel. Lastly, advanced 4D STEM-based[27] DPC imaging was used to obtain high-spatial-resolution images of the domain structures and states occurring during the transition. Taken together, these reveal new fundamental details of the mechanisms associated with the AF to FM phase transition at the highest spatial resolution, which were previously inaccessible.

## **Results**

Ordered $\alpha$'-FeRh alloy thin films were grown epitaxially on a clean (001) MgO substrate or (001) NiAl buffer layer on GaAs substrate by conventional DC magnetron sputter co-deposition, as described previously[29]. We studied two TEM specimens. Firstly, a cross-sectional FeRh TEM specimen was prepared from its bulk MgO substrate and transferred onto *in situ* heating electronic (e-) chip (DENSsolution Wildfire™) by $Ga^+$ focused ion beam (FIB) methods[24]. After FIB preparation, the cross-sectional sample was annealed at 600°C under vacuum in the TEM for 1 hour to recover any damage to the FeRh structure from $Ga^+$ implantation. Secondly, a planar FeRh TEM specimen was prepared through a process of HF-etching of AlAs, GaAs and NiAl buffer layers, as well as GaAs substrate[30], and subsequently transferred onto a Cu TEM grid for *in situ* heating using a Gatan heating TEM holder.

Figure 1 presents a cross-sectional view of the FeRh thin film grown on the MgO substrate, providing information on its thickness, chemical distribution, localised structure and interfacial magnetism with the MgO substrate. The high angle annular dark-field (HAADF) STEM image of Fig. 1a reveals the FeRh film to be grown with a uniform thickness of ~ 53 nm, whilst the electron-energy loss spectroscopy (EELS) chemical maps (Fig. 1b) acquired from the boxed region (red) in Fig. 1a display the elemental distribution of iron, rhodium and oxygen. The high resolution (HR) STEM image of Fig. 1c presents the localised CsCl crystal structure of the FeRh thin film along the ⟨110⟩ zone axis, where the alternating columns of Rh atoms appear brighter compared to the Fe atoms due to their higher atomic number (Fig. 1c, inset). Similarly, Fig. 1d displays the localised interface between the single crystalline FeRh and MgO substrate, revealing their well-matched orientation and confirming the epitaxial growth of the deposited FeRh. Segmented detector DPC imaging, presented in Fig. 1e, provides low temperature (30°C) magnetic information on the FeRh thin film. Within the FIB Pt protective layer, multi-coloured contrast is observed which is purely electrostatic in origin and arises due to it containing many grain boundaries. Within the FeRh layer, there is an absence of strong contrast, commensurate with it being in the AF state at this temperature. Strong contrast is observed between the FeRh layer and its interfaces. At the interface with the FIB Pt protective layer, this contrast relates to the sharp electron wave phase gradient which arises due to the two materials having different mean electrostatic potential (analogous to refractive indices in optics). This gives rise to a single green coloured band that indicates the strength and direction of this phase gradient which is purely electrostatic in origin. The situation at the interface between the FeRh/MgO substrate appears to be more complicated. As for the previously discussed interface, a change in the mean electrostatic potential between the two materials would be expected to contribute a single coloured band. However, here, both a strong red coloured band and a weaker green coloured band (indicated by an arrow) are observed. We ascribe the strong red band to being electrostatic in origin but the weaker green band to being associated with the existence of a finite magnetic moment over a region ~ 8 nm wide. This is explained in more detail in the Supplementary Information (Fig. S1-S3) and the Discussion section. In order to isolate the magnetic contrast induced by thermal effects during the magnetostructural transition, Fig 1e acts as a reference and is subtracted from DPC images in the temperature series presented in Figure 2. The steps for this subtraction process are also described in the Supplementary Information (Fig. S4).

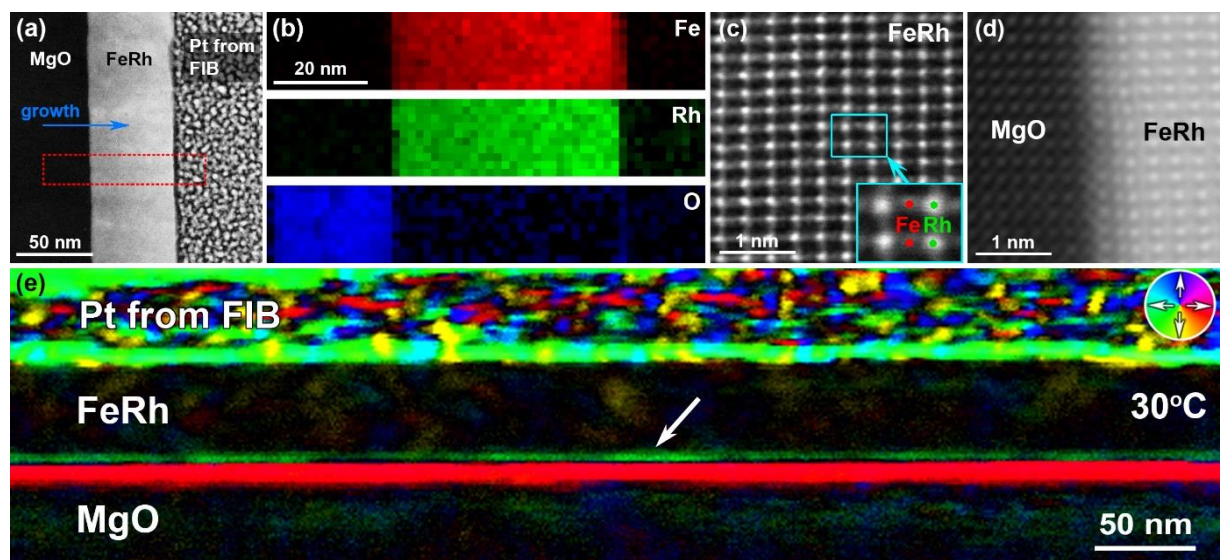

**Figure 1. Overview of the TEM cross-section of the FeRh thin film grown on a MgO substrate.** (a) HAADF-STEM image of the cross-sectional TEM lamella showing the FeRh thin film grown on the MgO substrate. (b) EELS chemical maps acquired from the box region in (a) displaying the elemental distribution of iron (red), rhodium (green) and oxygen (blue). (c,d) High-resolution HAADF-STEM images showing the localised structure of the (c) FeRh thin film with labelled atoms (inset); and (d) its interface with the MgO substrate. (e) Segmented detector DPC image of the FeRh thin film, revealing magnetic contrast at its interface with the MgO substrate (~ 8 nm green layer, arrowed). The direction of magnetization is depicted in the colour wheel (inset).

The segmented detector DPC images of Figure 2 present the magnetic domain evolution associated with the magnetostructural transition of the FeRh thin film as a function of temperature. Fig. 2a reveals nucleation of a small, green magnetic domain (~ 50 nm long, ~ 20 nm wide, labelled 'D'), with magnetisation directed from right to left, on the right side of the FeRh / MgO interface at 89°C. As the temperature is increased to 91°C the small domain is observed to act as a nucleation site for expansion to the left (Fig. 2b), and further growth at 93°C (Fig. 2c) and 103°C (Fig. 2d). Weak contrast associated with the domain's resulting dipolar magnetic field (yellow to blue) in Fig. 2a-d is observed in the MgO substrate as the magnetostructural transition proceeds. At 118°C, a large magnetic domain (red) is observed to form on the left hand-side of the FeRh thin film (Fig. 2e), with the magnetisation pointing from left to right. As the temperature is increased to 200°C, the two larger domains (red and green) are seen to be separated by a head-to-head transverse DW (blue, inset), like those seen in permalloy nanowires[14].

Figure 3 presents a planar view of the HF-etched FeRh thin film, providing details of its surface, morphology, chemistry, relative thickness and magnetism. The DF STEM image of Fig. 3a reveals the FeRh thin film to exhibit a non-uniform surface and morphology, with variations in contrast attributed to an inconsistent thickness and debris on the surface. The EDX chemical maps (Fig. 3b) acquired from the boxed region (red) in Fig. 3a display the elemental distribution of iron, rhodium, gallium and arsenic, revealing a relatively uniform distribution of rhodium and iron, larger gallium content on the left, along with concentrated areas of arsenic. The arsenic-rich areas coincide well with the surface debris in Fig. 3a and are expected to be caused by the HF-etching process. Fig. 3c presents a HAADF image of a large square area (~ 4 µm x ~ 4µm) and the SAED pattern (Fig. 4c, inset) confirms that the HF-etched FeRh film is single crystalline. Fig. 3d presents the thickness map acquired from the entire region of Fig. 3c, where the relative thickness ranges from ~ 85 nm at the red spots, artefacts attributed to grains lying on a crystallographic zone axis which scatter the electron beam away from HAADF detector, to a more uniform thickness of ~ 40 – 45 nm (average of ~ 43 nm). The DPC image of Fig. 3e reveals the presence of several magnetic domain structures (arrowed) when heated to 80°C, while most of the sample is in the AF state, represented by the black regions, where no magnetic deflections are detected.

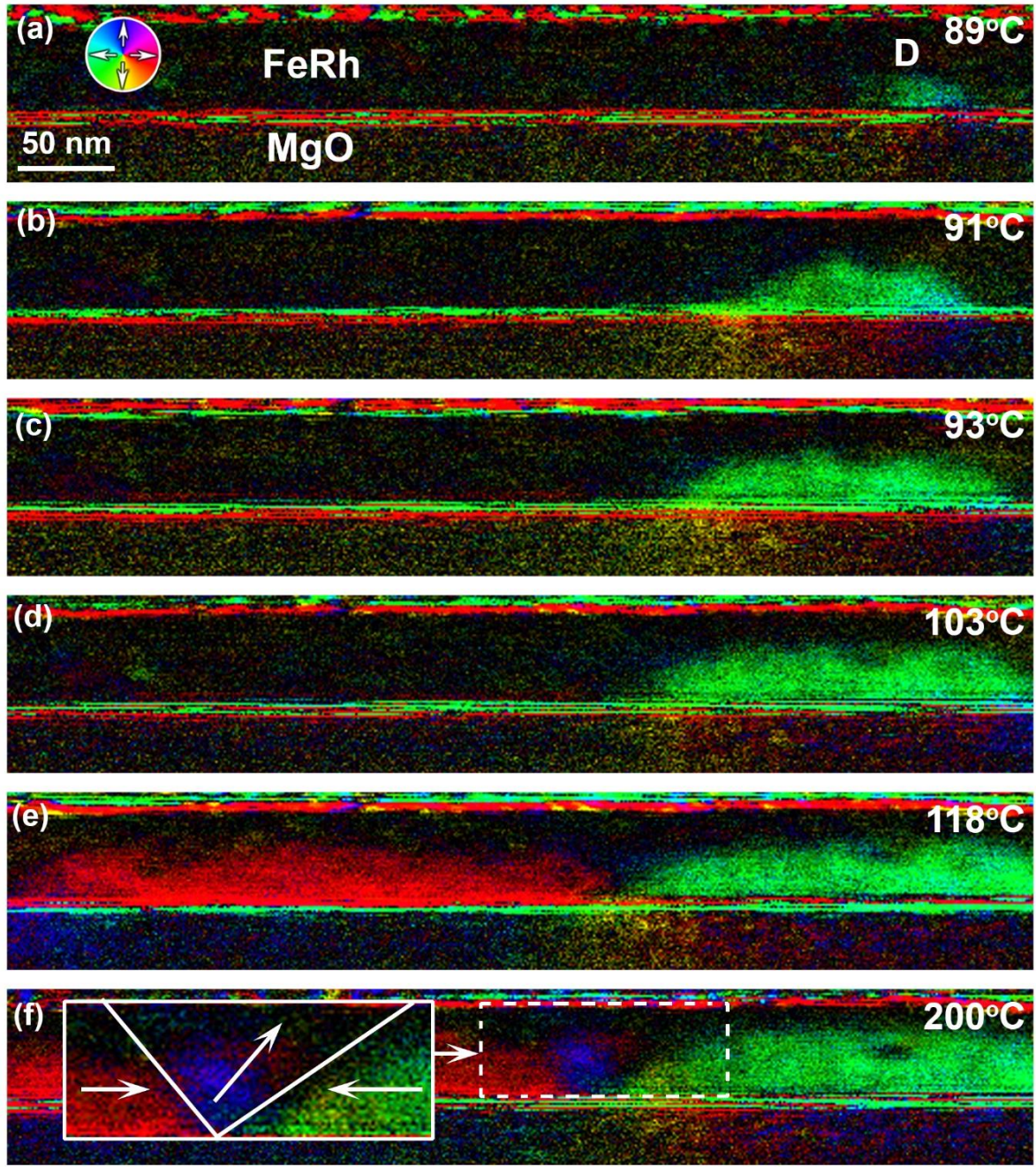

**Figure 2. Segmented detector DPC imaging of magnetic domain evolution in the FeRh thin film as a function of temperature.** (a-f) DPC imaging of the cross-sectional TEM sample of the FeRh thin film / MgO substrate during *in situ* heating to (a) 89°C; (b) 91°C; (c) 93°C; (d) 103°C; (e) 118°C; and (f) 200°C. The direction of magnetization is depicted in the colour wheel (a, inset) and is illustrated in the head-to-head transverse DW in (f) (blue, inset).

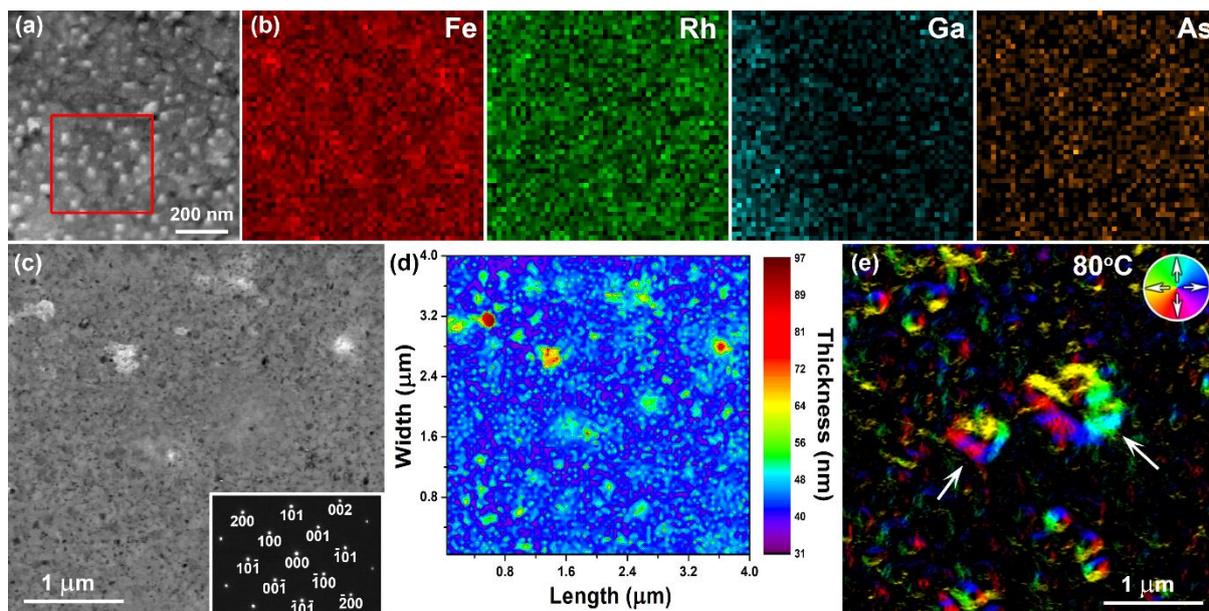

**Figure 3. Overview of the planar FeRh thin film after HF-etching of its AlAs, GaAs and NiAl buffer layers, as well as the GaAs substrate.** (a) DF STEM image of the HF-etched planar FeRh thin film. (b) EDX chemical maps acquired from the box region in (a) showing the distribution of iron, rhodium, gallium and arsenic. (c) HAADF STEM image showing several white spots that are attributed to grains lying on a zone axis which deflect the electron beam away from HAADF detector, with the SAED (inset, bottom right). (d) Thickness map calculated from the low-loss EEL spectrum acquired from the entire area of (c), with small regions of large thickness (red) coinciding with the white spots seen in (c). (e) DPC image of the same region in (c&d) during *in situ* heating, showing the presence of a magnetic domain structures at 80°C (arrowed).

Figure 4 provides direct visualisation of the thermomagnetic behaviour displayed by the HF-etched planar FeRh thin film using complementary techniques of Fresnel TEM imaging, small-angle electron scattering (both acquired sequentially from the same sample region) and 4D STEM DPC imaging (acquired from the same sample but from a different location). The relative scale of the 4D STEM scan region (256 x 256 pixels) to the Fresnel images is denoted in the bottom-right corner of Fig. 4a. The Fresnel TEM image of Fig. 4a, acquired at 75°C, shows a large area of the planar film (~ 15 μm in diameter) which includes two relatively strong contrast bend contour features (dark, near vertical bands) and Fresnel contrast that indicate the presence of small vortex structures[31] (arrowed). The corresponding SAES pattern (Fig. 4b) is acquired from the same area shown in Fig. 4a (obtained by alteration of the excitation of only the post-specimen lenses of the TEM) and reveals that the intensity of the electron beam is concentrated in the central spot, recording minimal dispersion from FM domains. This is supported by the associated 4D STEM DPC image acquired at 75°C (Fig. 4c), which documents the existence of vortex-like structures (~ 130 nm, white arrows) in close proximity to a more complex domain (centre of Fig. 4c). Nevertheless, the majority of the DPC image is covered by dark regions and provides complementary evidence indicating that the FeRh thin film is mostly in the AF state. After increasing the temperature to 87.5°C, the Fresnel TEM image (Fig. 4d) showed a significant increase in vortex state nucleation across the entire illuminated area, as well as formation of larger magnetic domains elongated along the orthogonal x-axis (arrowed) and slight variation in the bend contours. Fig. 4e reveals the accompanying development of the SAES pattern through the redistribution of intensity from the central spot to the outer concentric ring. Further insight is provided by the corresponding 4D STEM DPC image (Fig. 4f), revealing nucleation of additional FM vortex structures (arrowed, white) and their interconnection via small, uniformly magnetised 'string' domains (arrowed, black). Fig. 4g demonstrates that increasing the temperature further, to 90°C, further nucleation, and string domains (arrowed, black) are induced which coalesce into larger magnetic domains. The SAES pattern of Fig. 4h shows that most intensity is now located in the outer ring rather than the central spot, indicating that the film is in a mostly FM state with a full range of magnetic orientations present. This description of the state is supported by the 4D STEM DPC image of Fig. 4i. The Fresnel TEM image of the Fig. 4j shows that heating to 100°C promotes agglomeration into large, fully formed FM domains and reveals the bend contours to be straighter and more parallel compared to Fig. 4a. All the electron intensity in the corresponding SAES pattern (Fig. 4k) is segregated to the outer top-left arced ring pattern, which indicates the sample is in the fully FM state at 100°C, supported by the existence of large, fully-formed FM domains in the 4D STEM DPC image (Fig. 4l).

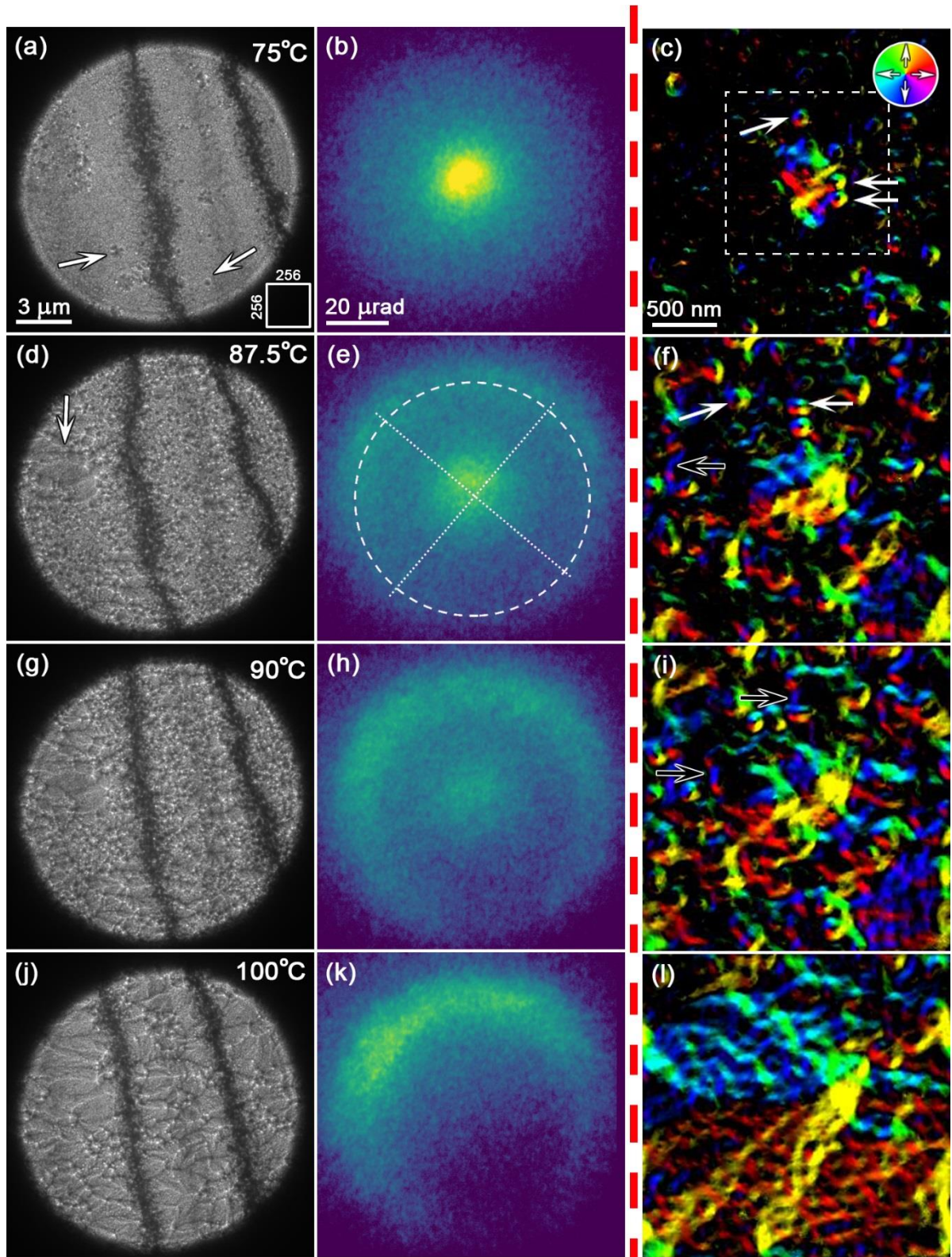

**Figure 4. Magnetic domain evolution in the planar FeRh thin film as a function of temperature during *in situ* heating to 100°C.** (a,d,g,j) Fresnel; (b,e,h,k) SAES; and (c,f,i,l) 4D STEM DPC images of a HF-etched FeRh thin film acquired at (a-c) 75°C; (d-f) 87.5°C; (g-i) 90°C; and (j-l) 100°C. The images demonstrate the different stages of the magnetostructural transition, including (a-c) nucleation of magnetic vortices (arrowed); (d-f) AF / FM phase co-existence; (g-i) domain coalescence; and (j-l) a fully FM state. The relative scale of the 256 x 256 4D STEM DPC images to the Fresnel images is denoted in the bottom-right left corner of Fig. 4a, while the associated direction of magnetization is depicted in the colour wheel (Fig. 4c, inset). The superimposed crosshairs in (e) are used to show the geometric centre of the SAES pattern.

In a similar fashion to Figure 4, Figure 5 charts the magnetostructural transition of the HF-etched planar FeRh thin film during stages of cooling from the FM state. The large FM domains in the Fresnel and 4D STEM DPC images of Fig. 5a and Fig. 5c, respectively, and outer arced ring pattern of the corresponding SAES pattern (Fig. 5b) suggests the FeRh is in the fully-FM state at 80°C, revealing hysteresis in the transition during cooling compared to heating. It is clear that the FM domains start to disintegrate at 70°C (Fig. 5d,f), as evidenced by electron intensity migrating towards the centre of the SAES pattern (Fig. 5e). The FM to AF transition proceeds further when cooled to 65°C, with almost all the large FM domains in Figs. 5g and 5i transforming into smaller vortex or 'string' states, and increased intensity in the central spot of the SAES pattern (Fig. 5h). The bend contours are observed to increasingly widen and ripple in Figs. 5g and 5j, with very few magnetic domains still present in the latter at 60°C. The majority of electron intensity in the SAES pattern acquired at 60°C (Fig. 5k) is concentrated in the central spot, with the associated 4D STEM DPC image of Fig. 5l displaying only a few vortex or string domain states.

Figure 6 presents quantitative analysis of selected SAES patterns acquired during heating through the AF to FM transition. The SAES pattern of Fig. 6a, at 85°C, contains two contributions: the central diffuse spot and the outer ring. For the central spot, the peak intensity was located at a deflection angle, $β_l$, of 6.4 μrad from the geometric pattern centre (denoted by crosshairs in Fig. 4e). The outer ring corresponds to fully FM regions possessing the complete range of in-plane orientation directions and its diameter is due to magnetic deflection of the electrons through a maximum angle $β_l$ of ~ 35 μrad. Fig. 6b displays the SAES pattern acquired at 140°C where the intensity has evolved, becoming concentrated in an arc of restricted angular range on the outer ring (top-left side) possessing the same deflection angle amplitude, $β_l$ of ~ 35 μrad. This deflection angle, $β_l$, can be related to the saturated magnetic induction for the FeRh thin film using the equation:

$$B_s = \frac{β_l h}{eλt} \qquad (1)$$

where $B_s$ is the saturation induction, $t$ is the thickness of the magnetic thin film (measured from the EELS low-loss thickness map), $h$ is Planck's constant, $e$ is the magnitude of electronic charge and $λ$ is the electron wavelength. Using equation 1, the saturation induction can be calculated from the SAES pattern for the fully-FM state, assuming the film is uniformly magnetised throughout its thickness. The $β_l$ of 35 μrad in the SAES pattern of Fig. 6b corresponds to a saturation induction of 1.34 T ± 0.15 T, where the uncertainty is attributed to the full width half maximum spread of intensity in the arced outer ring of the SAES pattern. This value is in very good agreement with the bulk saturation magnetisation measurements from this HF-etched FeRh thin film in its FM state of 1060 emu/cm$^3$, or 1.33 T[26]. Considering the AF to FM transition is of first order, without intermediate stages of reduced $B_s$ from the FM phase, it is assumed the $β_l$ of 6.4 μrad in Fig. 6a corresponds to FM regions of reduced thickness. Again, using equation 1 and the $B_s$ = 1.33, the $β_l$ of 6.4 μrad is calculated to correspond to FM regions with a thickness of 7.9 ± 0.2 nm, which is commensurate with the ~ 8 nm wide FM region observed in Fig. 1e, and will also be explained in more detail in the Discussion section.

The 4D STEM DPC images of Figure 7 provides a localised examination of domain evolution from the white-boxed region in Fig. 4c, as a function of temperature. Fig. 7a shows that the central complex domain at 70°C comprises a vortex (arrowed, white) attached to a blue 'string' domain flowing from right to left and separated from the rest of the domain by an elongated AF region, before wrapping around on the left-hand side. As the temperature is increased to 75°C (Fig. 7b), the elongated AF region disappears and the two vortices (arrowed) on the right-hand side become more prominent. Further heating to 80°C promotes the merging of these two vortices (arrowed) into the larger central domain (Fig. 7c). This domain is extended to the left at 85°C in Fig. 7d through growth of a large vortex-like state (arrowed, white) and 'string' domain (arrowed, black). Increasing the temperature to 87.5°C promotes separation of the oppositely magnetised top and bottom of the left-hand side of this magnetic structure (blue and yellow, respectively) by a black DW (arrowed). At 90°C, the dark DW disintegrates into a cross-tie DW (arrowed), and the central domain is observed to integrate with domain structures on all sides.

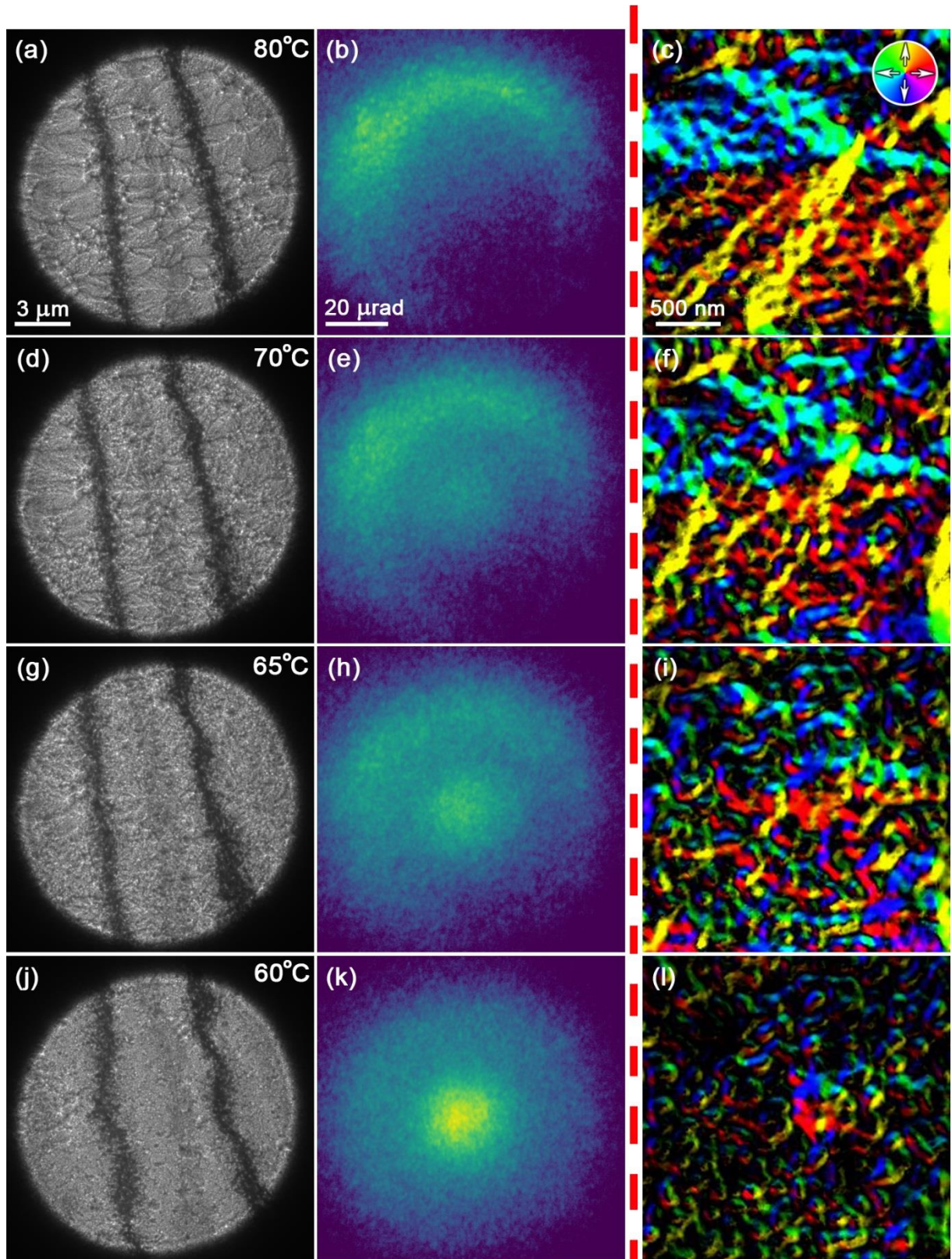

**Figure 5. Magnetic domain evolution in the planar FeRh thin film as a function of temperature during *in situ* cooling from 80°C.** (a,d,g,j) Fresnel; (b,e,h,k) SAES; and (c,f,i,l) 4D STEM DPC images of a HF-etched FeRh thin film acquired at (a-c) 80°C; (d-f) 70°C; (g-i) 65°C; and (j-l) 60°C. The images chart the reversal of the magnetostructural transition from (a-c) large domains of the FM state; (d-f) disintegration of FM domains; (g-i) AF / FM phase co-existence with small FM domains in the form of vortices and 'strings'; and (j-l) small vortices. The direction of magnetization in the DPC images is depicted in the colour wheel (Fig. 5c, inset).

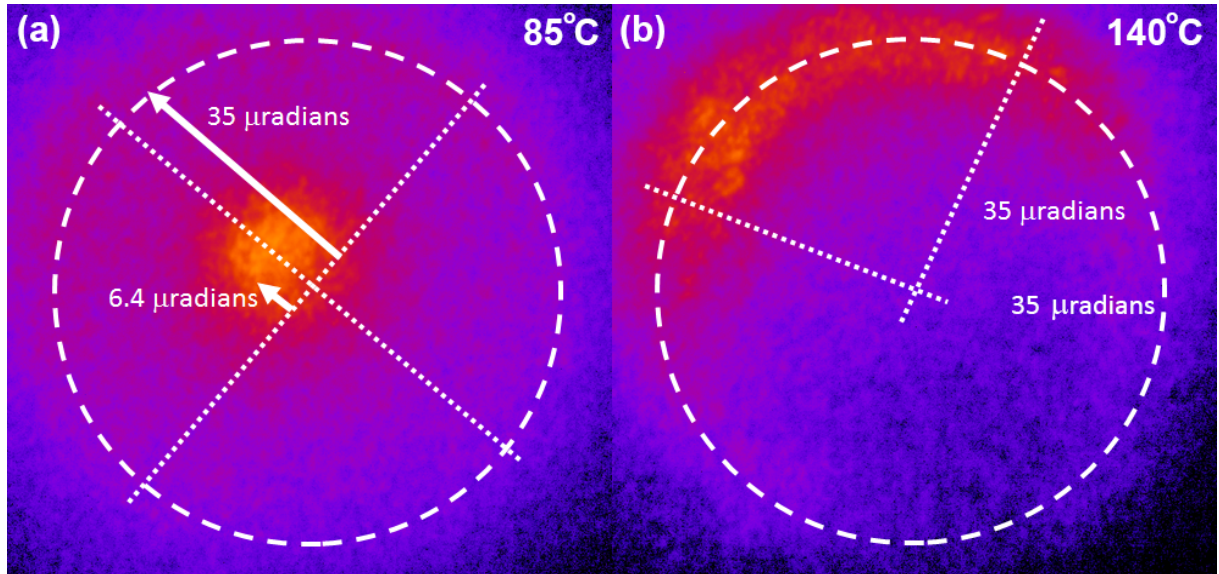

**Figure 6. Quantitative analysis of selected SAES patterns.** (a) SAES pattern acquired at 85°C, showing the highest intensity corresponding to a 6.4 µrad deflection from the central optic axis (denoted by cross-hairs), along with a diffuse ring of intensity at a deflection of ~ 35 µrad. (b) SAES pattern acquired at 140°C showing the highest intensity at a deflection of ~ 35 µrad towards the top left of the SAES pattern.

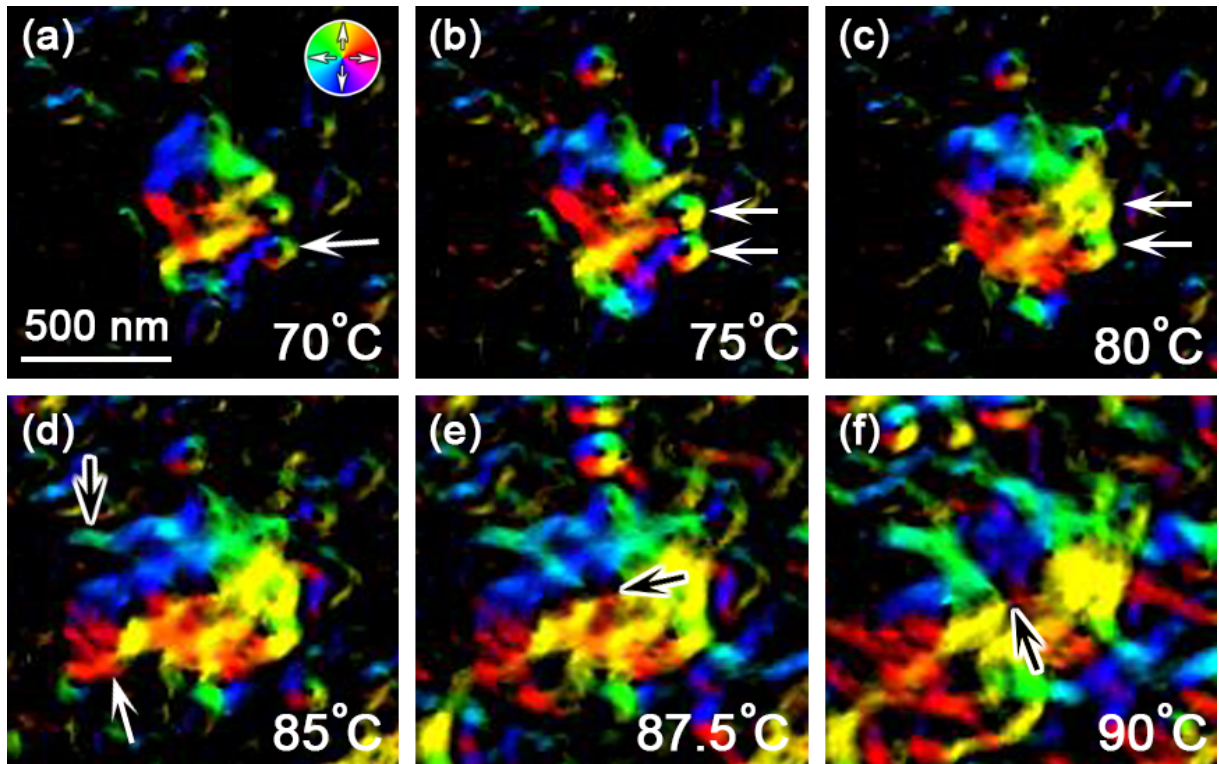

**Figure 7. Localised magnetic domain evolution during *in situ* heating.** (a-f) 4D STEM DPC images of the HF-etched FeRh thin film acquired at (a) 70°C; (b) 75°C; (c) 80°C; (d) 85°C; (e) 87.5°C; and (f) 90°C. The images reveal the subtle localised growth dynamics of FM domains during the AF to FM phase transition. The direction of magnetization in the DPC images is depicted in the colour wheel (Fig. 7a, inset).

**Discussion**

This multi-mode combined Lorentz microscopy investigation has provided fundamental, localised insight into the magnetostructural transition in FeRh thin films as a function of temperature. The DPC image of Fig. 1e revealed magnetic contrast in the vicinity of the interface with the MgO substrate at 30°C. The existence of such an interfacial moment has been previously detected indirectly and is attributed to interfacial strain induced within the surface atomic layers of the FeRh, due to the slight mismatch between the FeRh and MgO unit cells, and its thickness (6 – 8 nm) is consistent with that reported[32]. It is also suggested that interfacial oxidation can form a FeO-like layer that enhances the spin polarisation of this small FM surface interfacial barrier[33]. The DPC image series of Figure 2 recorded the AF to FM transition during heating from 89°C to 200°C in the cross-sectional FeRh sample. The first small FM domain (~ 50 nm wide, ~ 20 nm high) nucleated at 89°C on the right-hand side of the FeRh / MgO interface, which is consistent with previous reports of inhomogeneous spatial distribution of the transition along the growth direction, where the transition initiates at the substrate interface[25]. This domain grows both laterally and upwards into the cross-section with temperature until 118°C, at which point another domain rapidly nucleates and grows on the left-hand side. These domains are separated by an asymmetric transverse DW[34], likely due to the shape anisotropy of the cross-sectional FeRh sample, as they are not an energetically favourable configuration for planar FM films. Nevertheless, these domains can be compared to the uniformly-magnetised 'string' domains observed in the planar FeRh sample, as the thinner cross-sectional FeRh TEM lamella would not accommodate the small vortex states (~ 130 nm). Hence, this study goes beyond previous work on FeRh thin film cross-sections[25] by showing evolution of multiple domains separated by DWs, without the need to magnetically saturate the sample, e.g. for electron holography, which can provide a directional bias or memory, even in the AF state. Further, the domain evolution and formation of DWs in the FeRh cross-section is consistent with that observed in the planar FeRh film (Figs. 3&4).

In addition to the small FM signal at room temperature in the cross-sectional TEM sample (Fig. 1e), the SAES patterns also reveal a small FM signal detected from the planar FeRh film. It is considered that preferential HF-etching of the more reactive Fe atoms can result in a Fe-deficient and Rh-terminated surface, where symmetry breaking stabilizes a surface FM layer at room temperature[35]. This FM layer is believed to be responsible for the dispersed central spot in the SAES pattern acquired at 75°C (Fig. 4b) and provides a directional bias in terms of domain evolution. This is demonstrated by the 6.4 μrad offset from the optic axis (Fig. 6a) and arced outer ring favoured to the top left of Fig. 6b, as well as elongation of the magnetic domains in the Fresnel images. The 6.4 μrad offset corresponds to a ~ 8 nm thick surface FM layer, which is consistent with the ~ 6 – 8 nm interfacial FM region in the cross-sectional TEM sample. Hence, this study provides both visual and quantitative evidence of surface FM layers, as well as insight into initial stages of the domain evolution during the AF to FM transition in the planar FeRh film, the details of which will now be discussed in detail.

The 4D STEM DPC images revealed that the thermally-induced nucleation of vortex structures (~ 130 nm in diameter) act as initial or seed FM domain states in an AF matrix, where the reduction of magnetostatic energy at the AF / FM boundary during the formation of the vortex state is consistent with the edges of a FM nanostructure patterned in a non-magnetic film[23]. These vortex seeds clearly exhibit continuously flowing magnetization and no DWs, providing an unambiguous interpretation compared to previous XMCD results where a four-domain pattern separated by DWs and a vortex-state would appear identical[8]. The localised expansion of the crystal lattice associated with the FM domain formation is likely to induce strain in adjacent AF regions and make them more susceptible to transition to the FM state. This is due to the film not being constrained to a substrate and hence freer to expand, as demonstrated by the variation of bend contours observed in the Fresnel images. For this reason, isolated single domains are rarely observed at this high spatial resolution due to the ease of their transition to the more energetically favoured vortex state, as compared to substrate constrained FeRh films analysed by surface magnetic imaging techniques like XMCD[8] and SEMPA[9]. Hence, the initial stages progress through nucleation of vortex states in the close vicinity of FM vortices or 'string' domains spreading out from them (Fig. 4f & Fig. 7), where they are then seen to agglomerate and be consumed by the nearby domains with increasing temperature. In the case of the directional growth of string domains, crystal expansion is expected at the string-front prompting further directionally favoured growth, whilst it is considered regions parallel to the string domain may undergo momentary compression which hinders their transition and preserves dark AF regions between adjacent parallel string domains, as observed at the bottom of the central domain in Fig. 7a. At this stage, the inherently multi-directional nature of vortices and 'string' domains deflects the transmitted electron beam in all directions and is in good agreement with the concentric ring of intensity in the SAES pattern of Fig. 4e. As more FM vortices nucleate, the AF regions decrease in size and FM domains expand through agglomeration (Fig. 4g,i). This stage is considered to comprise small AF regions and all three main types of FM domain: vortices; 'strings'; and larger elongated domains which exhibit a preferred directionality due to their biased interaction with the FM surface layer. Once the AF regions fully transition to the FM state with increasing temperature, the remaining FM vortices and 'strings' lower their energy configurations through coalescence with the larger domains. The fully-FM state will also undergo long-range energy minimisation by merging of large grains through DW annihilation, as well as the associated formation of cross-tie DWs, seen in Fig. 7f.

It is clear from both the Fresnel and 4D STEM DPC imaging that the FM domain disintegration during the FM to AF transition behaves differently to the FM domain evolution described above. As FM regions become AF, there is no net localised magnetic moment and hence no need to minimise their energy into vortex structures like the FM states. However, in a similar fashion to the FM domains, the AF regions act as nucleation sites for AF region growth which disrupts the long-range ordering of the large FM domains. The high prevalence of the 'string' domains observed in Fig. 5f,i is considered to be due to the retention of directionality during the short-range disintegration of the larger FM domains. This is supported by the associated SAES patterns (Fig. 5e,h) exhibiting fewer concentric outer rings, favouring the top-left side, compared to heating (Fig. 4e,h). As the AF regions grow and 'strings' are the dominant FM domains, they are less likely to relax into vortices since their domain width is already < 100 nm. Further, the prominent, complex domain in the centre of Fig. 5l suggests that the first FM domains formed are also likely to be the last to disintegrate during cooling, confirming that local structural defects or thickness variations act as both favourable domain nucleation and final annihilation sites.

This comprehensive TEM study has revealed directly the AF / FM phase transition in nanoscale FeRh thin films as a function of temperature. We have provided both visual and quantitative evidence of surface FM layers in the cross-sectional and planar FeRh samples, respectively, which was previously inaccessible. Complementary segmented detector and 4D STEM DPC imaging displayed the intricate details of the evolution and dissipation of the FM domains at an unambiguous level of detail. The imaging was suitably complemented by SAES, providing quantitative measurements of integrated induction matching the bulk magnetic measurements. As a significant step forward in understanding the FeRh magnetic transition, this high-spatial-resolution magnetic imaging provides an explicit route to analysis of the more localised and complex thermally-activated PB motion in gradient-doped FeRh thin films, as well as current-induced spin injection[36].

## Methods
*Electron microscopy.* All the imaging, diffraction and spectroscopy described in this paper were carried out on a JEOL Atomic Resolution Microscope (JEM-ARM200F) TEM, operating at 200 kV[23]. This microscope is equipped with a cold field emission gun and a CEOS (Corrected Electron Optical Systems GmbH) probe corrector for STEM imaging. Conventional and HR STEM imaging were performed on cross-sectional and planar TEM samples of the FeRh films / substrates, whilst SAED acquired in TEM mode provided structural information. Both EDX and EELS provided chemical analysis of the samples. The sample thickness was determined by the spectrum imaging technique in STEM mode[37], whereby low-loss EELS spectra acquired from each pixel were used to calculate values of $t/\lambda$. These calculations were performed using the Digital Micrograph™ software package. The mean free path, $\lambda$, was determined from the density of equi-atomic FeRh and the TEM beam conditions, *i.e.*, accelerating voltage, convergence and divergence angles, *etc.* and substituted into $t/\lambda$ to calculate the relative thickness repetition, with a standard deviation of 6%. The magnetic structure of the FeRh films was visualised using segmented and 4D STEM DPC in Lorentz mode under low-magnetic field conditions. DPC imaging was carried out with the HR objective lens pole piece switched off, with the samples positioned in the low-strength remanent field of the lens (~ 150 Oe out of plane). An 8-segment silicon photodiode array detector (supplied by DebenUK Ltd) was used for the segmented DPC imaging. The signal from the detectors was converted and amplified using the "Superfast" amplifier (Andrew Armit Designs). The 8 detected signals were acquired, mixed and displayed via four Gatan DigiscanII units. For 4D STEM DPC imaging, the central diffraction disk was recorded at each scan position by a Medipix3 pixelated detector, whilst using a 50µm condenser aperture and camera length of 800cm. The cross-correlation disk deflection registration method was used to reconstruct the 4D STEM DPC images[27]. In addition, combining DPC with *in situ* heating using the DENSsolution e-chips (up to 200°C) or Gatan heating holder (up to 110°C) allowed for direct access to the thermomagnetic behaviour of the DWs and magnetostructural transition within the FeRh films. SAES patterns were acquired in TEM mode using a 30-micrometre diameter condenser aperture which yielded parallel illumination of a circular region of the sample, 15 microns in diameter. A custom post specimen lens excitation series was used to either obtain Fresnel images or SAES patterns with 460m camera length at each temperature step. Switching between the imaging / SAES mode was performed by only adjusting the strength of the first lens immediately after the main imaging lens.

**Acknowledgements**
The authors thank the EPSRC (grants EP/M019020/1 & EP/M018504/1) and Diamond Light Source for funding.

**Author contributions statement**
T.P.A. and D.M designed and carried out the experiments; C.H.M., D.M., S.M. and T. M. conceived the experiment; J.M. and R.T. synthesised and provided the samples; Y.L assisted with the experimental work; T.P.A, D.M., and S. M. analysed the results. T.P.A. led the writing of the paper and all authors reviewed the manuscript.

**Additional Information**
**Competing Interests:** The authors declare no competing financial interests.

**Data and materials availability**: All data needed to evaluate the conclusions of this study can be found the at the following link <link will be inserted once the manuscript has been accepted>.

# Direct visualisation of the magnetostructural phase transition in nano-scale FeRh thin films using differential phase constrast imaging

Trevor P. Almeida[1]*, Damien McGrouther[1], Rowan Temple[2], Jamie Massey[2], Yue Li[1], Thomas Moore[2], Christopher H. Marrows[2], Stephen McVitie[1]

[1] SUPA, School of Physics and Astronomy, University of Glasgow, Glasgow, G12 8QQ, UK.
[2] School of Physics and Astronomy, University of Leeds, Leeds LS2 9JT, UK.

## Supplementary information

*Corresponding author:
Tel: +44 (0) 141 330 4712
Email: trevor.almeida@glasgow.ac.uk

**This PDF file includes:**
S1: Calculation of phase profiles across the FeRh / MgO interface at 20°C.
S2: Isolation of magnetic contrast in FeRh thin film cross-section temperature series.

## S1 Calculation of phase profiles across the FeRh / MgO interface at 20°C

In order to understand the experimentally acquired DPC images of the cross-sectional FeRh sample, the electrostatic potential and magnetic contributions to the phase profiles were calculated. The total phase shift experienced by the transmitted electron beam is sensitive to the electrostatic potential and the in-plane component of the magnetic induction in the specimen, as summarised by:

$$\phi(x,y) = \phi_e + \phi_m$$

The volume average of the electrostatic potential, or mean inner potential (MIP), $V_0$, of MgO used was found directly from a previous study[1], whilst the values of FeRh were determined from modelled electrostatic potential values of Fe and Rh, as summarised in Table S1. The MIP of Fe and Rh were calculated using Thomas-Fermi-Dirac (TFD) or Thomas-Fermi (TF) statistical models of a compressed atom[2], and the average values taken. The mean of the MIP of both Fe and Rh were then used to infer a MIP value for equi-atomic FeRh.

|         | MgO  | Fe          |           |           | Rh          |         |           | FeRh      |
|---------|------|-------------|-----------|-----------|-------------|---------|-----------|-----------|
|         |      | $V_0^{TFD}$ | $V_0^{TF}$ | Avg. $V_0$ | $V_0^{TFD}$ | $V_0$ TF | Avg. $V_0$ | Avg. $V_0$ |
| MIP (V) | 13.0 | 19.4        | 21.7      | 20.5      | 25.0        | 27.6    | 26.3      | 23.4      |

**Table S1.** Values of MIP used for calculation of phase profiles across the FeRh / MgO interface.

The MIP contribution to the phase profile was calculated using the following equation:

$$\phi_e(x,y) = C_E V(x,y) t(x,y) \quad (1)$$

where the thickness ($t$) is 20 nm, and $C_E = 7.29 * 10^6$ V$^{-1}$m$^{-1}$ for 200 kV electrons. The MIP contributions to the phase in uniform sections of MgO and FeRh are:

ϕ (MgO) = 1.90 radians
ϕ (FeRh) = 3.41 radians

Figure S1 displays the MIP contribution to the phase across the width of the FeRh / MgO interface (measured as ~ 8 nm in Fig. S4b, red boxed region).

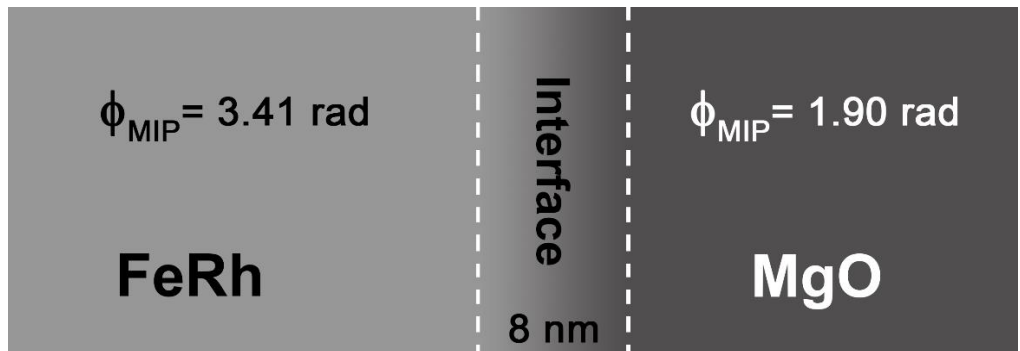

**Figure S1.** Schematic of the MIP contribution to the phase across the FeRh / MgO interface.

The magnetic contribution to the phase was calculated using the bulk measurement of saturation induction ($B_s$) = 1.33 T and $t$ = 12 nm, assuming 4 nm non-magnetic surface layers due to irradiation damage and amorphization:

$$\beta_l = \frac{e\lambda B_s t}{h} = \sim 9.7 \text{ μrad} \quad (2)$$

Regions of uniform magnetisation would lead to phase gradients with strength:

$$\dot{\phi} = \frac{2\pi}{\lambda}\beta = 0.024 \ rad/\ nm^{-1} \quad (3)$$

Using the 8 nm length of the FM region measured from the experimental DPC image, then a magnetic phase change of: 8 * 0.024 = 0.194 radians, or from -0.97 radians to +0.97 radians, occurs between the FeRh / MgO interface and 8nm into the FeRh thin film, represented graphically in Fig. S2.

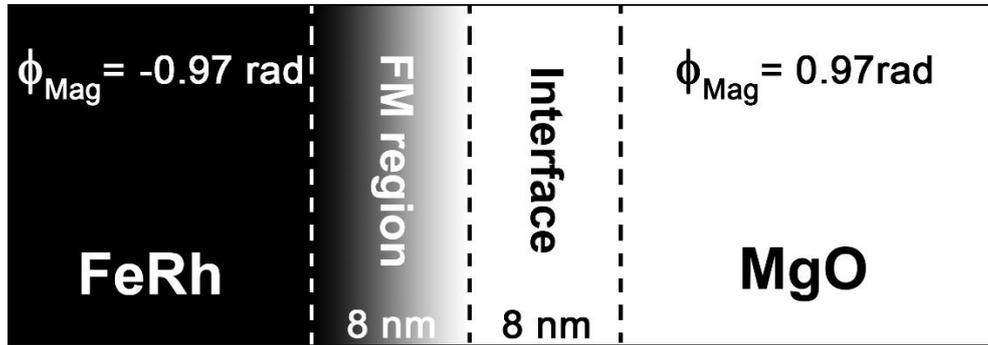

**Figure S2.** Schematic of the magnetic contribution to the phase across the FeRh / MgO interface, with an adjacent 8 nm FM region in the FeRh film.

The individual calculated $\phi_{MIP}$ and $\phi_{Mag}$ profiles across the width of the FeRh / MgO interface and 8nm FM region, as well as their sum $\phi_{Total}$, are plotted in Fig. S3a. In Fig. S3b, the derivative of $\phi_{Total}$, or phase gradient, $\phi'_{Total}$, is used in equation 2 to calculate a modelled deflection angle ($\beta_l$), which is superimposed with the average line profile of $\beta_l$ directly measured from the difference signals of the associated DPC images, as shown in Fig. S4b (red boxed region). Fig. S3b clearly demonstrates the similarity of the profiles across the FeRh / MgO interface with an adjacent 8 nm FM region in the FeRh thin film, reinforcing our suggestion of a FM layer due to interfacial strain with MgO substrate.

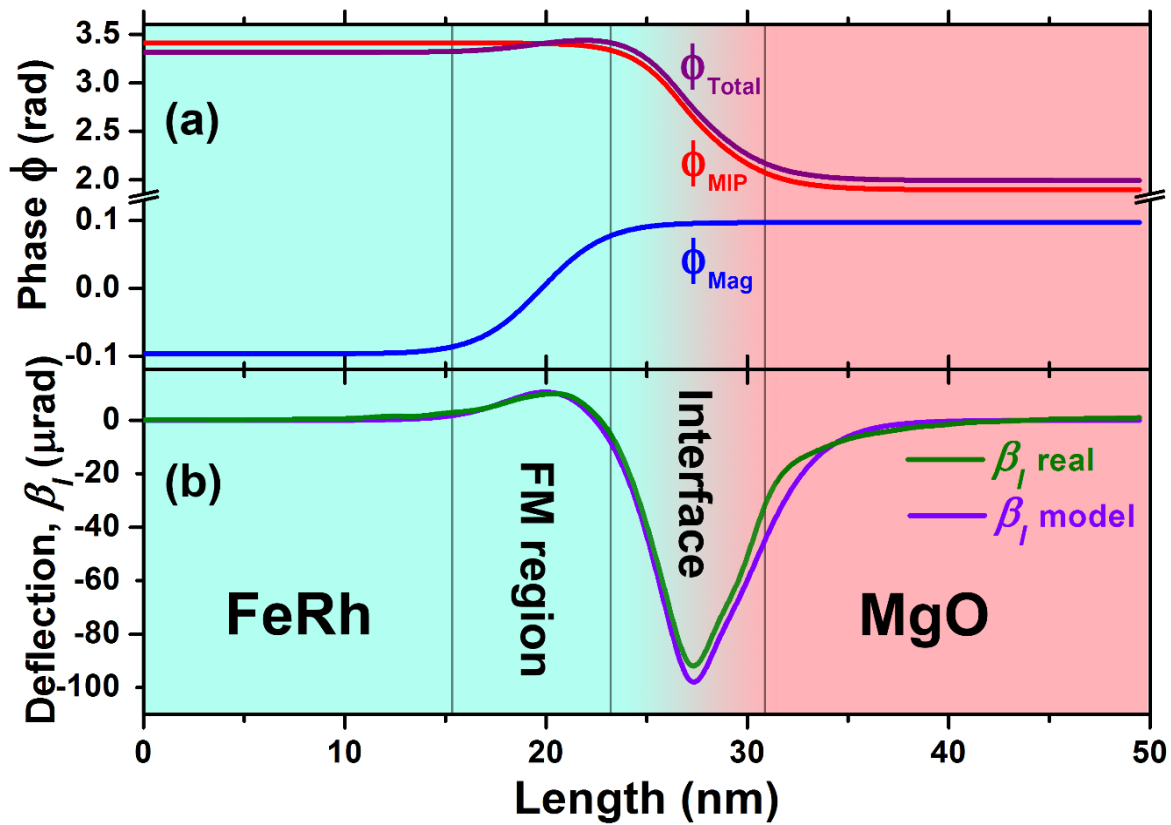

**Figure S3.** (a) Profiles of the magnetic, electrostatic and combined contribution to the phase across the FeRh / MgO interface, with an adjacent 8 nm FM region in the FeRh film. (b) Profiles of the modelled and measured deflection angles ($\beta_l$) across the FeRh / MgO interface and 8 nm FM region, demonstrating similar profiles.

## S2 Isolation of magnetic contrast in FeRh thin film cross-section temperature series

In order to investigate the domain evolution in the FeRh cross-section as a function of temperature, the room temperature DPC images acted as a benchmark and were subtracted from the temperature series to isolate the magnetic contrast, as summarised in Figure S4. For example, the difference signals acquired at 20°C (Fig. s4a,b) were aligned and subtracted from those acquired at 93°C (Fig. S4c,d), resulting in difference signal more representative of the change in magnetic signal in the FeRh thin film (Fig. S4e,f). The difference signals are combined to illustrate the direction of magnetic induction using colour (Fig. S4g). The colour image is then combined with the normalised magnitude image of the difference signals (Fig. S4h) to create a magnetic induction map of FeRh in a state of AF / FM coexistence at 93°C (Fig. S4i).

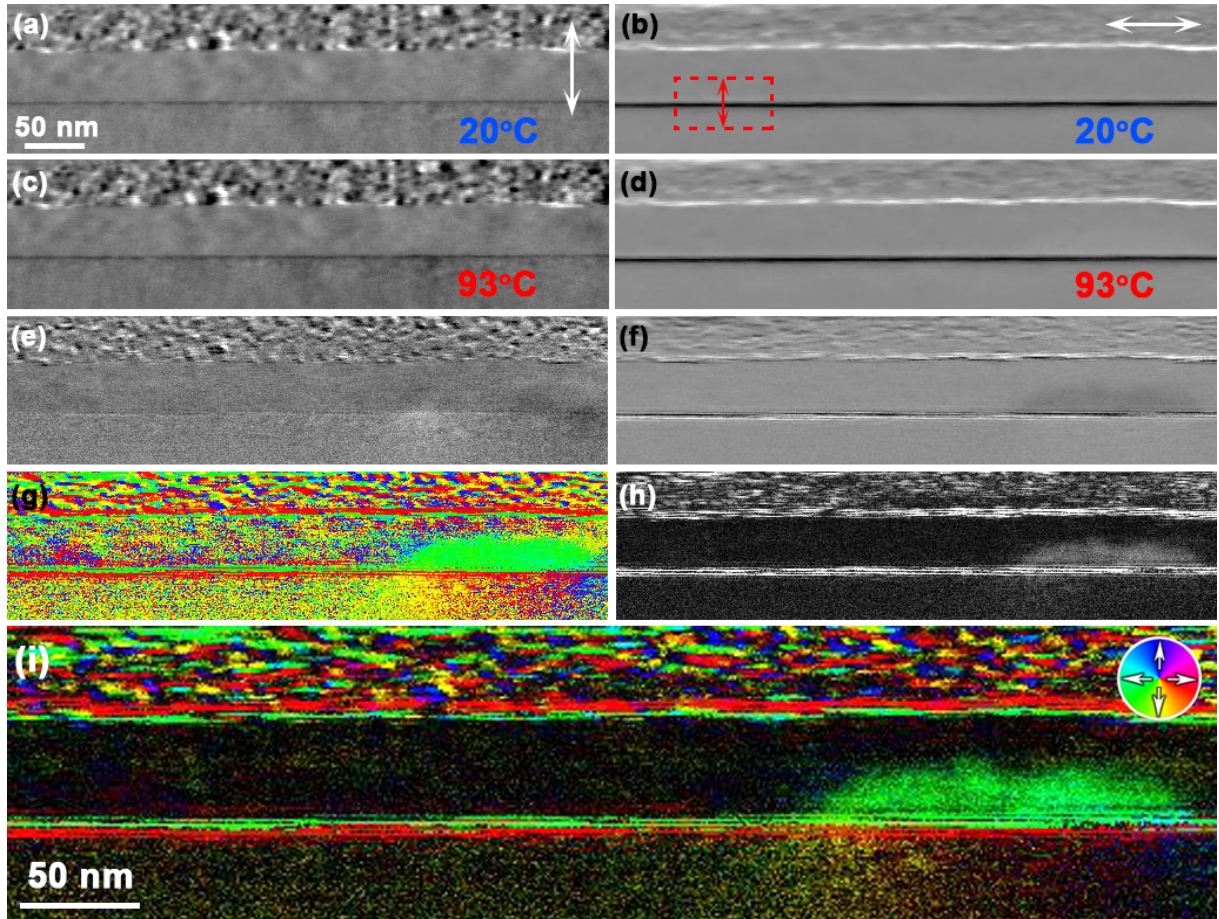

**Figure S4.** Difference signals of the DPC images acquired at (a,b) 20°C; (c,d) 93°C; and (e,f) the result of their subtraction. These DPC images are sensitive to the component of magnetic induction indicated by the double-headed white arrows in (a) for the left panel; and (b) for the right panel. (g) Colour image of the difference signals showing the direction of magnetic induction. (h) Normalised magnitude image of the difference signals. (i) Magnetic induction map showing a small FM domain (right) at the FeRh / MgO interface at 93°C.